\documentclass[reprint,aps,prd,amsmath,showpacs]{revtex4-1}
\usepackage{graphicx}
\usepackage[linkcolor=blue,citecolor=blue]{hyperref}
\usepackage{bm}
\usepackage{float}
\usepackage{amssymb}
\usepackage{latexsym}
\usepackage{scrextend}
\hypersetup{colorlinks=true}

\newcommand{\mm}{\mu\bar{\mu}}
\begin{document}
\title{Some radiative corrections to the hyperfine splitting of true muonium: Two-photon-exchange contributions}
\author{Yao Ji}
\email{yao.ji@physik.uni-regensburg.de}
\affiliation{Institut f\"ur Theoretische Physik, Universit\"at Regensburg, Regensburg 93040, Germany}
\author{Henry Lamm}
\email{hlamm@umd.edu}
\affiliation{Department of Physics, University of Maryland, College Park, MD 20742}
\date{\today}

\begin{abstract}
We consider a set of radiative contributions from one-loop lepton vacuum polarization to the hyperfine splitting of true muonium.  Improving previous results, we obtain values for the electron-loop coefficients and extract their leading dependence upon $\beta=m_\mu/m_e$.  The coefficients are $C^{\mu}_{20,1{\rm VPCT}}=0.734931603\beta$, $C^{\mu}_{20,1{\rm VPCA}}=0.551198702\beta$, and $C^\mu_{20,1{\rm VPT}}=0.419149715\beta$.  The mass-independent contribution from three one-loop vacuum polarization is $C_{40,{\rm VPX}}=\frac{1}{127575}\left(-5255-2016\pi^2+51840\zeta(3)\right)$.  Contributions from $\tau$ up to $\mathcal{O}(m_\mu\alpha^8)$ are calculated.
\end{abstract}
\maketitle

True muonium is the yet unidentified $(\mm)$ bound state with lifetimes on the order of ps~\cite{Brodsky:2009gx}.  QED dominates the 
characteristics of true muonium, while QCD and Electroweak effects appear at 
$\mathcal{O}(m_\mu\alpha^5)$~\cite{Lamm:2016vtf} and $\mathcal{O}(m_\mu\alpha^7)$~\cite{PhysRevD.91.073008} respectively.  The need to discover and study true muonium is motivated by the existing discrepancies in muon 
physics~\cite{PhysRevD.73.072003,Antognini:1900ns,Aaij:2014ora,Aaij:2015yra,
Pohl1:2016xoo}.  Both new physics models~\cite{TuckerSmith:2010ra,Jaeckel:2010xx,Batell:2011qq,
Barger:2011mt,Karshenboim:2010cm,Karshenboim:2010cg,Karshenboim:2010cj,
Karshenboim:2010ck,Karshenboim:2011dx,Karshenboim:2014tka,Carlson:2012pc,
Carlson:2015poa,Izaguirre:2014cza,Kopp:2014tsa,Martens:2016zzx,Liu:2016qwd,
Onofrio:2013fea,Wang:2013fma,Gomes:2014kaa,Brax:2014vva,Lamm:2015gka,
Lamm:2016jim} and systematic errors in the experiments have been proposed to resolve these 
discrepancies.  Other works have suggested a more subtle understanding of known physics is required~\cite{Jentschura:2014ila,Miller:2015yga,Stryker:2015ika,Burgess:2016lal,Burgess:2016ddi,Burgess:2016zbp}.  True muonium can produce competitive constraints on all these solutions if the standard model predictions are known to the 100 MHz level, corresponding to 
$\mathcal{O}(m_\mu\alpha^7)$~\cite{Lamm:2015fia}.  Today, the Heavy Photon Search (HPS)~\cite{Celentano:2014wya} experiment is searching for true muonium~\cite{Banburski:2012tk}, and DImeson Relativistic Atom Complex (DIRAC)~\cite{Benelli:2012bw} has discussed a search during an upgraded run~\cite{dirac}.  In both situations, the true muonium will be relativistic, necessitating consideration to the effect of boost on wave functions and production rates~\cite{Lamm:2013oga,Lamm:2016ong,Lamm:2016djr}.  

In this work, we will focus on improving the theoretical prediction for the hyperfine splitting (hfs).  To review, the expression for the hfs corrections to true muonium from QED can 
be written
\begin{align}
\label{eq:eng}
 \Delta E_{\rm hfs}=m_\mu\alpha^4\bigg[&C_{0}+C_{1}\frac{\alpha}{\pi}+C_{21}\alpha^2\ln\left(\frac{
1}{\alpha}\right)+C_{20}\left(\frac{\alpha}{\pi}\right)^2\nonumber\\
 &+C_{32}\frac{\alpha^3}{\pi}\ln^2\left(\frac{1}{\alpha}\right)+C_{31}\frac{
\alpha^3}{\pi}\ln\left(\frac{1}{\alpha}\right)\nonumber\\
 &+C_{30}\left(\frac{\alpha}{\pi}\right)^3+\cdots\bigg],
\end{align}
where $C_{ij}$ indicate the coefficient of the term proportional to 
$(\alpha)^i\ln^j(1/\alpha)$ and for $i=0,1$ the second index is dropped.  $C_{ij}$ include any dependence on mass scales other than $m_\mu$ (e.g $m_e,m_\pi,m_\tau,$...).  The coefficients of single-flavor QED bound 
states, used in positronium, are known up to $\mathcal{O}(m_e\alpha^6)$. Partial results have been computed for $\mathcal{O}(m_e\alpha^7)$ and are an active research area in light of upcoming experiments (For an updated review of 
the coefficients see \cite{Adkins:2014dva,PhysRevA.94.032507}).  The exchange 
$m_e\rightarrow m_\mu$ translates these results to true muonium.

True muonium receives further contributions that are typically neglected in positronium.  Most importantly, the existence of the lighter 
electron allows for large vacuum loop contributions.  The relative smallness of 
$m_\tau/m_\mu\approx 17$ and $m_{\pi}/m_{\mu}\approx1.3$ produce non-negligible contributions as well.  Of 
these true muonium specific contributions, denoted by $C_{ij}^\mu$, 
only a few terms are known.  

The hadronic contribution to the annihilation channel, $C^\mu_{1,\rm hvp}=-0.04874(9)$, has been recently computed using the experimental Drell ratio $R(s)$~\cite{Lamm:2016vtf}.  The leptonic-loops in the two-photon annihilation channel coefficient
$C^\mu_{20,2\gamma}=-2.031092873$ is known exactly~\cite{PhysRevA.94.032507}, and the electron loop in three-photon 
annihilation has been calculated numerically $C_{30,3\gamma}^\mu=-5.86510(20)$~\cite{Adkins:2015jia}. The leading-order contributions from $Z$-bosons has also been computed~\cite{PhysRevD.91.073008}.

Ref.~\cite{Jentschura:1997tv} presented calculations for the correction to the $n=1,2$ $S$-states from electron vacuum polarization in the Coulomb line with an additional transverse photon (VPCT), vacuum polarization in the Coulomb line with an annihilation photon (VPCA), and the vacuum polarization in a transverse photon (VPT).  With $C_{1,\rm hvp}$ now known to a higher precision, the uncertainty from finite precision in these contributions, 200 MHZ, is as large as the unknown higher-order contributions and therefore must be removed.  These results would naively scale as $\mathcal{O}(m_\mu\alpha^6)$, but the electron loops modify this scaling to $\mathcal{O}(m_\mu\frac{m_\mu}{m_e}\alpha^6)\approx\mathcal{O}(m_\mu\alpha^5)$.  This large enhancement over naive $\alpha$ scaling led~\cite{Jentschura:1997tv} to assign this contribution to $\mathcal{O}(m_\mu\alpha^5)$, finding it to be for the ground state
\begin{align}
 \tilde{C}_{1,e}^\mu=&\frac{\alpha}{\pi}\frac{m_\mu}{m_e}\left[C_{20,1{\rm VPCT}}^\mu+C_{20,1{\rm VPCA}}^\mu+C_{20,1{\rm VPT}}^\mu\right]\nonumber\\=&0.353+0.265+0.201=0.819.
\end{align}
Together, these contributions predict $\Delta E^{1s}_{\rm 
hfs}=42329429(16)_{\rm had}(200)_{1,e}(700)_{\rm miss}\text{ MHz}$ where the first uncertainty estimate 
is from hadronic experimental uncertainties, the second from the finite precision of $\tilde{C}_{1,e}$ computed in~\cite{Jentschura:1997tv}, and the final is from 
uncalculated $\mathcal{O}(m_\mu\alpha^6)$ contributions.
 
In this work, we will instead assign these contributions to their correct $\alpha^6$ scaling and compute them to higher precision.  Along the way we will also correct errors in the literature and in the case of the VPT derive an analytic expression.  This completely reduces the uncertainty from finite precision.  We compute these contributions for the dimensionless variable $\beta=\frac{m_\mu}{m_{e}}$.

Additionally, we use the scattering approximation to compute the full radiative contribution at $\mathcal{O}(m_\mu\alpha^n)$ from $\mu$ and $\tau$ loops in the two-photon exchange channel where $n=6,7,8$.  We reproduce the $\mu$-loop values of $C_{20,{\rm VPX}}=\frac{5}{9}$~\cite{Sapirstein:1983xr} and $C_{30,{\rm VPX}}=\frac{1}{3}\left(\frac{6\pi^2}{35}-\frac{8}{9}\right)$~\cite{Eides:2014nga,Eides:2015nla} (note that $E_F^{Ps}=m\alpha^4/3$ in the references) and compute for the first time the $C_{40,{\rm VPX}}=\frac{1}{127575}\left(-5255-2016\pi^2+51840\zeta(3)\right)$ term.  For the $\tau-$loops, numerical results are obtained.  

\section{VPC-T and VPC-A}
The vacuum polarization insertion into the Coulomb line with an additional transverse photon (VPCT) and with an annihilation photon (VPCA) are related by their relative contribution to the hyperfine splitting (4/7 and 3/7, respectively) and therefore we only need compute one.  Following~\cite{Jentschura:1997tv}, we compute VPCT.  The contribution is given by
\begin{align}
 \Delta E_{n{\rm VPCT}}=&\frac{4}{7}\frac{\alpha}{\pi}\frac{E_F}{n^3}\left[\frac{\Delta\psi_{nS}(0)}{\psi_{nS}(0)}\right]\nonumber\\
		     =&\frac{m_\mu\alpha^6}{\pi^2}C^\mu_{20,n{\rm VPCT}}\, ,
\end{align}
where $n$ is the energy level and we emphasize that the coefficients have an $n$ dependence, and $E_F$ is the Fermi energy.  $\Delta\psi_{nS}(0)/\psi_{nS}(0)$ is the correction to the wave function, given by
\begin{equation}
 \frac{\Delta\psi_{nS}(0)}{\psi_{nS}(0)}=2\int_{\Omega}\mathrm{d}^3r\,\bar{G}_{nS}(E_{nS};0,r)V_U(r)\psi_{nS}(\bm{r}).
\end{equation}
where $\Omega$ indicates an integration over all space.  The Uehling potential is given by
\begin{equation}
 V_U(r)=-\frac{\alpha^2}{\pi r}\int_0^1\mathrm{d}v\frac{v^2(1-v^2/3)}{1-v^2}e^{-\lambda r},
\end{equation}
where $\lambda=2m_{e}/\sqrt{1-v^2}$. The reduced Coulomb Green's function, $\bar{G}_{nS}(E_{nS};0,r)$, can be expressed in closed form for $S$ states~\cite{karshenboim1995logarithmic}.  For the cases of $n=1,2$, these formulae are
\begin{align}
 \bar{G}_{1S}(E_{1S};0,r)=&\frac{\alpha m_r^2}{2\pi}\frac{e^{-z_1/2}}{z_1}\nonumber\\&\times\left[2z_1(\ln z_1+\gamma)+z_1^2-5z_1-2\right],
\end{align}
and
\begin{align}
 \bar{G}_{2S}(&E_{2S};0,r)=-\frac{\alpha m_r^2}{8\pi}\frac{e^{-z_2/2}}{z_2}\nonumber\\&\times\bigg[4z_2(z_2-2)(\ln z_2+\gamma)+z_2^3-13z_2^2+6z_2+4\bigg],
\end{align}
where $z_n=2\alpha m_r r/n$ and $\gamma$ is Euler's constant.  The integrals over $\bm{r}$ can be done analytically and the remaining $v$ integral for the $1S$ state is given by
\begin{align}
\label{eq:vpct}
 C^{\mu}&_{20,1{\rm VPCT}}=\pi\beta\int\limits^1_0\mathrm{d}v\frac{v^2(3-v^2)}{9\theta(2+x\theta)^3}\nonumber\\&\times\left[8+14x+3x^2\theta^2-2x\theta(2+x\theta)\log\left(\frac{x\theta}{2+x\theta}\right)\right],
\end{align}
where $\theta=\sqrt{1-v^2}$, and $x=\alpha\beta=\frac{\alpha m_{\mu}}{m_e}$.  An integral expression in terms of only $x$ and $v$ can be found for $2S$ as well, but is omitted for length.  
Numerically integrating these expressions, we find $C^{\mu}_{20,1{\rm VPCT}}=0.734931603\beta$, and $C^{\mu}_{20,2{\rm VPCT}}=0.635841279\beta$.  Multiplying by $3/4$ we obtain  $C^{\mu}_{20,1{\rm VPCA}}=0.551198702\beta$, and $C^{\mu}_{20,2{\rm VPCA}}=0.476880959\beta$.  Multiplying these results by $\alpha/\pi$, we find agreement with the values of ~\cite{Jentschura:1997tv} with improved precision.  We point out there is an error in their Eq.~(18) as printed because when we recomputed this expression numerically, we find it equivalent to $C^{\mu}_{20,1{\rm VPCT}}=1.16\beta$, much larger than the correct value of $0.734931603\beta$.  The contributions to the hfs are
\begin{align}
 \Delta E_{1{\rm VPC}}=&\frac{m_\mu\alpha^6}{\pi^2}\left[C^{\mu}_{20,1{\rm VPCT}}+C^{\mu}_{20,1{\rm VPCA}}\right]\nonumber\\=&103948.8793\text{ MHz}\, ,
\end{align}
\begin{align}
 \Delta E_{2{\rm VPC}}=&\frac{m_\mu\alpha^6}{\pi^2}\left[C^{\mu}_{20,2{\rm VPCT}}+C^{\mu}_{20,2{\rm VPCA}}\right]\nonumber\\=&89933.5235\text{ MHz}\, .
\end{align}

\section{VPT}
The contribution of one-loop electron vacuum polarization in a transverse photon is given by the interaction between the magnetic field induced by the Uehling potential and the muons.  The general expression for these terms to a given $nS$ state is  
\begin{align}
\Delta &E_{n{\rm VPT}}=\nonumber\\& {\frac{8\pi}{3m_\mu^2}}\int^\infty_0dr\,r^2\left(-\frac{\partial}{\partial r}\big|\psi_{nS}(r)\big|^2\right)\left(\frac{\partial}{\partial r}V_U(r)\right).
\end{align}
In deriving their expressions, it is clear that \cite{Jentschura:1997tv} omitted the nontrivial boundary term to write the expressions $-(\vec{\nabla}|\psi|^2)\cdot(\vec{\nabla}V_U)$ as $|\psi|^2\vec{\nabla}^2V_U$ in their intermediate steps, while their final results are correct.  The integrals can be reexpressed into our standard notation as 
\begin{align}
 \Delta E_{n{\rm VPT}}&=\frac{\alpha^6}{\pi^2}C^{\mu}_{20,n{\rm VPT}}\, ,
\end{align}
where for $n=1$ and $x<2$, we have derived an analytic expression:
\begin{widetext}
\begin{align}
C^{\mu}_{20,1{\rm VPT}}=\frac{\pi\beta}{27x^4}\left\{x(24+x^2)+3\frac{16-2x^2+x^4}{\sqrt{4-x^2}}\left[\pi-2\text{arctan}\left(\frac{x}{\sqrt{4-x^2}}\right)\right]-24\pi\right\}\, ,
\end{align}
similarly for $n=2$ and $x<4$, we have:
\begin{align}
 C^{\mu}_{20,2{\rm VPT}}=&\frac{7\pi\beta}{5184x^4(16-x^2)^{5/2}}\bigg\{x\sqrt{16-x^2}(49152-5632x^2-232x^4+11x^6)\nonumber\\&+6\left[\pi-2\text{arctan}\left(\frac{x}{\sqrt{16-x^2}}\right)\right](65536-10240x^2+1056x^4-40x^6+x^8)-384\pi(16-x^2)^{5/2}\bigg\}\, .
\end{align}
\end{widetext}
For true muonium, these coefficients are $C^\mu_{20,1{\rm VPT}}=0.419149715\beta$ and $C^\mu_{20,2{\rm VPT}}=0.031410727\beta$, which agree with Eq.~(25) of Ref.~\cite{Jentschura:1997tv} but are exact up to uncertainties in the physical constants.  These correspond to
\begin{align}
 \Delta E_{1{\rm VPT}}=33876.9275\text{ MHz}\, ,
\end{align}
\begin{align}
 \Delta E_{2{\rm VPT}}=21.0480555\text{ MHz}\, .
\end{align}

\section{Using the Scattering Approximation}
\begin{figure}
 \includegraphics[width=0.40\linewidth]{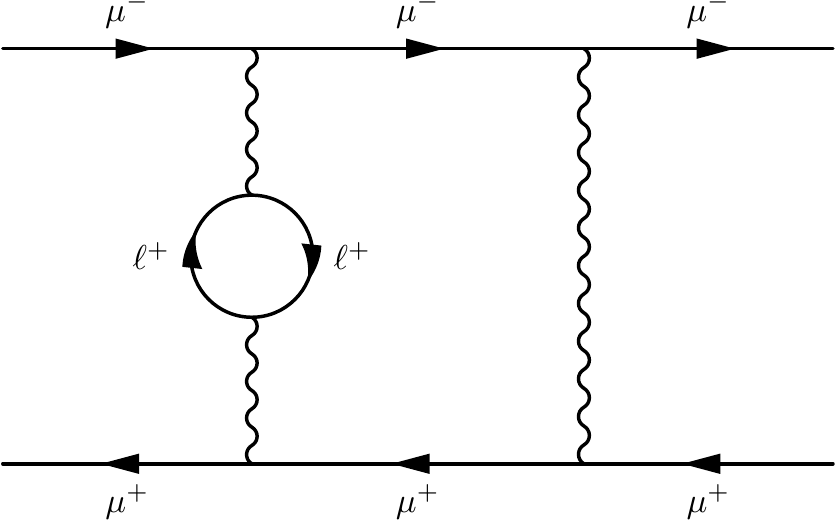}\\\vspace{.2cm}
 \includegraphics[width=0.40\linewidth]{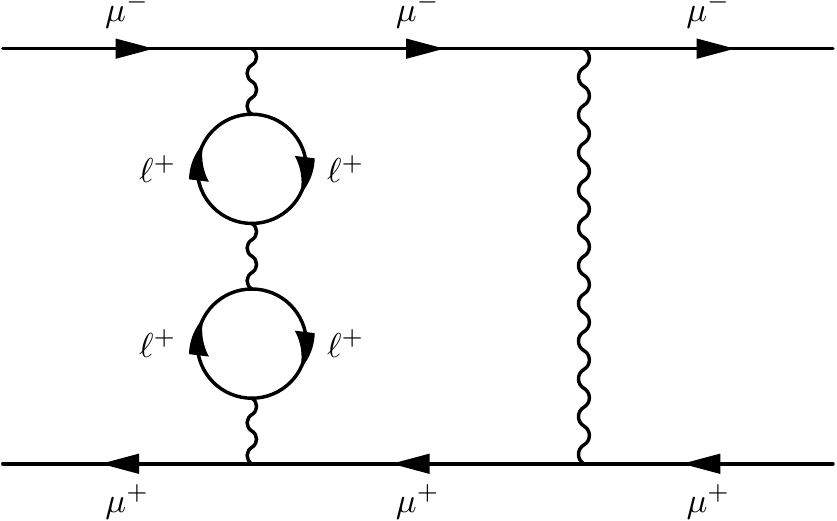}\hspace{.3cm}
 \includegraphics[width=0.40\linewidth]{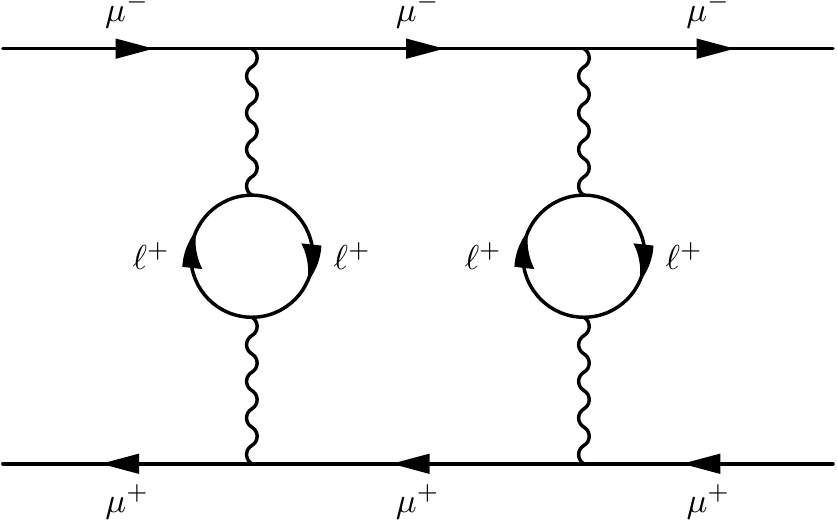}\vspace{.2cm}
 \includegraphics[width=0.40\linewidth]{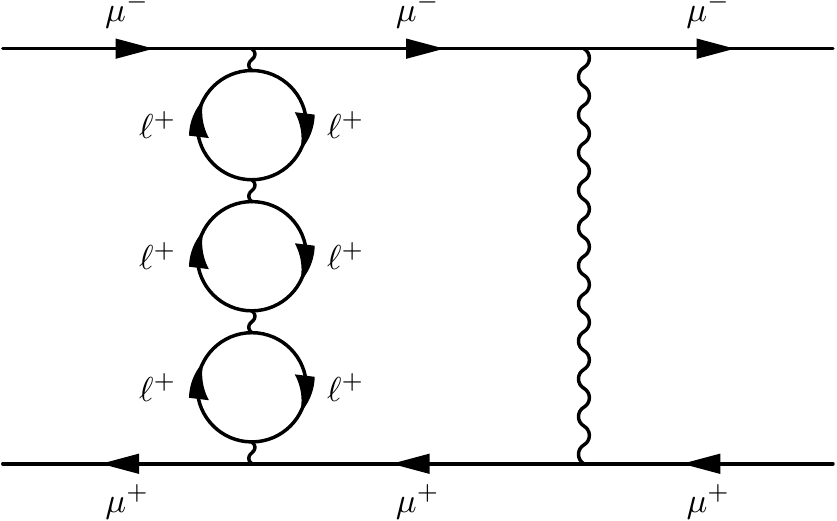}\hspace{.3cm}
 \includegraphics[width=0.40\linewidth]{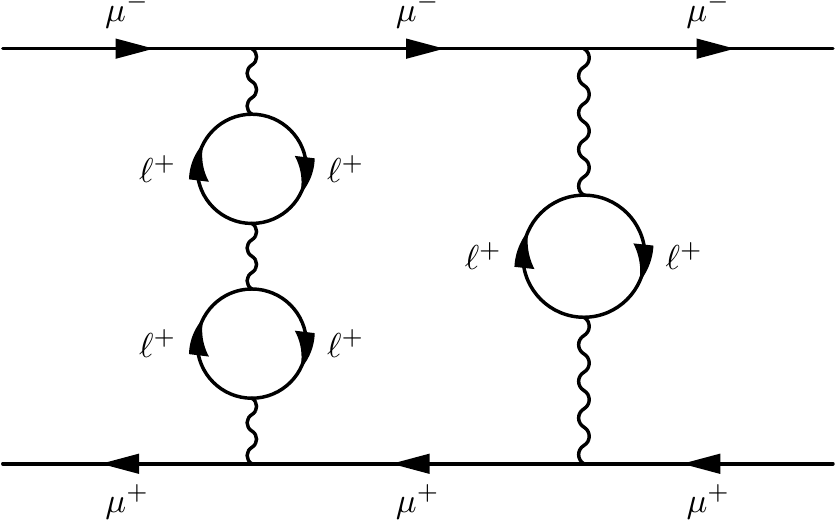}
 \caption{\label{fig:1}Radiative loop corrections from heavy leptonic loops to the two photon exchange graph considered in this work.}
\end{figure}
We wish to consider the heavy lepton contributions from Fig.~\ref{fig:1} to the $1S$ state within the scattering approximation.  This approximation multiplies $|\psi(0)|^2$ by the loop integral with the on-mass-shell fermions. These results are valid when the loop momenta are much larger than the momenta of the bound state, $\mathcal{O}(m_\mu\alpha)$. Insertions of vacuum polarization loops to the skeleton diagram effectively makes exchanged momenta $\mathcal{O}(m_{j})$ where $j$ is the loop particles.    The infrared-divergent skeleton diagram of the two-photon exchange obtained in the scattering approximation which requires $m_\ell'\gg\alpha m_\ell$.  It is found to have a simple form~\cite{Eides:1994mm,Eides:2014nga,Eides:2015nla}:
\begin{equation}
\label{eq:skel}
 \Delta E_{{\rm VPX}}=\frac{m\alpha^5}{\pi}\left[\frac{1}{3}\int_{0}^{\infty}\mathrm{d}q^2f_p(q)\right],
\end{equation}
where
\begin{equation}
 f_p(q)=\frac{16+2q^2+q^4-q^3\sqrt{q^2+4}}{4q^3\sqrt{q^2+4}}\, .
\end{equation}
While this integral itself is infrared-divergent, all radiative insertions into the expression will render it convergent.

The one-loop polarization insertion given in momentum space is
\begin{equation}
 \frac{\alpha}{\pi}I_{1,\beta}(q)=\frac{\alpha}{\pi}\int^1_0\mathrm{d}v\frac{v^2\left(1-\frac{v^2}{3}\right)}{(1-v^2)q^2+\frac{4}{\beta}}.
\end{equation}
  The contribution of $n$ one-loop vacuum polarization insertions is obtained from Eq.(\ref{eq:skel}) with $\left(\frac{\alpha}{\pi}q^2I_{1,\beta}(q)\right)^n$ and the appropriate combinatoric factor.  The integral over $v$ can be performed analytically, and the resulting $q^2$ integral is
\begin{widetext}
\begin{align}
\label{eq:q}
\Delta E_{1{\rm VPX}}=&\frac{m\alpha^5}{\pi}\left(\frac{\alpha}{\pi}\right)^{n-1}\left[\frac{n}{3}\int_{0}^{\infty}\mathrm{d}q^2f_p(q)\left(\frac{(12-5q^2\beta)+6(q^2\beta-2)\sqrt{\frac{q^2\beta+4}{q^2\beta}}\text{arctanh}\left(\sqrt{\frac{q^2\beta}{q^2\beta+4}}\right)}{9q^2\beta}\right)^{n-1} \right]\notag\\
 =&\frac{m\alpha^5}{\pi}\left(\frac{\alpha}{\pi}\right)^{n-1} C_{n0,{\rm VPX}}\, ,
\end{align}
where the factor of $n$ reflects the combinatorics.
 \end{widetext}
In the case of $\beta=1$, we can analytically solve this equation for $n=2,3,4$.  Our results agree with previous calculations for $C_{20,{\rm VPX}}=5/9$~\cite{Sapirstein:1983xr} and $C_{30,{\rm VPX}}=\frac{1}{3}\left(\frac{6\pi^2}{35}-\frac{8}{9}\right)$~\cite{Eides:2015nla,Eides:2014nga}.  These terms have already been included in the hfs of true muonium.  While experimental precision even for positronium has not reach reached this level, we found $C_{40,{\rm VPX}}=\frac{1}{127575}\left(-5255-2016\pi^2+51840\zeta(3)\right)$.  The contribution from this term is $0.000614$ MHz in true muonium and $0.00297$ kHz in positronium.  

For values of $\beta<1$, we have been unable to find an analytic expression, and have instead numerically integrated Eq.~\ref{eq:q} for $n=2,3,4$.  The results for physically relevant cases are found in Tab.~\ref{tab:1}.  The contribution from $\tau$-loop terms in true muonium is $2.225430+0.001023+~0.000002=2.226455$ MHz.

\begin{table*}
 \caption{\label{tab:1}Coefficients from two-photon exchange from one-loop leptonic vacuum polarization for physical values of $\beta_{ij}$ where $i$ is the valence lepton, and $j$ is the lepton in the loop.}
 \begin{center}
 \begin{tabular}{l c c c}
 \hline\hline
  $\beta$&$C^{\beta_{ij}}_{20,{\rm VPX}}$&$C^{\beta_{ij}}_{30,{\rm VPX}}$&$C^{\beta_{ij}}_{40,{\rm VPX}}$\\
  \hline
  $\beta_{e\tau}=2.9\times10^{-4}$&   $3.0946311(1)\times10^{-7}$&$2.6420938(1)\times10^{-8}$&$2.5388718(1)\times10^{-8}$\\
  $\beta_{e\mu}=4.8\times10^{-3}$&    $6.1110193(1)\times10^{-5}$&$7.4715685(1)\times10^{-6}$&$7.1799093(1)\times10^{-6}$\\
  $\beta_{\mu \tau}=5.9\times10^{-2}$&$5.6932879(1)\times10^{-3}$&$1.1262411(1)\times10^{-3}$&$1.0851353(1)\times10^{-3}$\\
  $\beta_{\ell\ell }=1$&$\frac{5}{9}$&$\frac{1}{3}\left(\frac{6\pi^2}{35}-\frac{8}{9}\right)$&$\frac{1}{127575}\left(-5255-2016\pi^2+51840\zeta(3)\right)$\\
  \hline\hline
 \end{tabular}
\end{center}
\end{table*}

We have fit a large number of numerical values to find the $\beta$ scaling for $n=2,3,4$. The coefficients are increasingly well-approximated by $C_{n0,{\rm VPX}}\propto\beta$ with increasing $n$.  The $C_{n0,{\rm VPX}}\propto\beta$ behavior at large $n$ agrees with naive scaling expectations inside loops, and can be used to guide estimates of higher order corrections.
\begin{table}
 \caption{\label{tab:2}Power law dependence extracted from fits to numerically integrated results for $C_{n0,{\rm VPX}}\sim\beta^{\kappa}$.}
 \begin{center}
 \begin{tabular}{l c}
 \hline\hline
  $C_{n0,{\rm VPX}}$&$\kappa$\\
  \hline
  $n=2$&1.082(1)\\
  $n=3$&1.013(1)\\
  $n=4$&1.0031(4)\\
  \hline\hline
 \end{tabular}
\end{center}
\end{table}
\section{Conclusion}

In conclusion, we have computed several critical corrections to the hyperfine splitting of true muonium necessary to reach 100 MHz precision.  The recalculation of the VPCT, VPCA, and VPT diagrams has reduced the error budget by 200 MHz by removing the error from finite precision and in the case of VPT has yielded analytic results.  We have further computed $C_{n0,{\rm VPX}}$ for $n=2,3,4$ for values of $\beta$ that are physically relevant.  The current theoretical prediction is $\Delta E^{1s}_{\rm 
hfs}=42329435(16)_{\rm had}(700)_{\rm miss}\text{ MHz}$.  In light of this work, the largest uncertainty arises from uncalculated $\mathcal{O}(m_\mu\alpha^6)$ electron loops.

\begin{acknowledgments}
HL would like to express thanks to Michael Eides for pointing out errors in an earlier draft of this work.  HL is supported by the National Science Foundation under Grant Nos. PHY-1068286 and PHY-1403891 and by the U.S. Department of Energy under Contract No. DE-FG02-93ER-40762.  YJ acknowledges the Deutsche Forschungsgemeinschaft for support under grant BR 2021/7-1.
\end{acknowledgments}

\bibliography{wise}
\end{document}